\documentclass[prb,aps,twocolumn,showpacs,amsmath,amssymb]{revtex4-1}
\usepackage{graphicx}
\usepackage{dcolumn}
\usepackage{bm}
\usepackage{epsfig}
\usepackage{color}
\definecolor{grey}{gray}{0.75}
\usepackage{amssymb}
\usepackage[french]{babel}
\usepackage[T1]{fontenc}
\usepackage{ulem}
\usepackage{graphicx}
\usepackage{epsf}

\begin{document}
\title{Temperature driven Vanadium clusterization and band gap enlargement in the layered misfit compound (LaS)$_{1.196}$VS$_2$ }
\author{V. Ta Phuoc}
\affiliation{GREMAN, CNRS UMR 7347, Universit\'e F. Rabelais, UFR Sciences,
Parc de Grandmont, 37200 Tours, France}
\author{V. Brouet}
\affiliation{Laboratoire de Physique des Solides. CNRS UMR8502\\Universit\'{e} Paris-Sud - Bat.510. 91405 Orsay Cedex - France}
\author{B. Corraze, E. Janod, L. Cario}
\affiliation{Institut des Mat\'{e}riaux Jean Rouxel. CNRS UMR 6502 CNRS\\Universit\'{e} de Nantes. 2, rue de la Houssini\`ere. BP 32229. 44322 Nantes - France }
\begin{abstract}
Intriguing properties of the misfit layered chalcogenide (LaS)$_{1.196}$VS$_2$ crystals were investigated by transport, optical measurements, angle-resolved photoemission (ARPES) and x-ray diffraction. Although no clear anomaly is found in transport properties as a function of temperature, a large spectral weight transfer, up to at least 1 eV, is observed by both optical and photoemission spectroscopies. ARPES reveals that a nearly filled band with negative curvature, close enough from the Fermi level at 300K to produce metallic-like behaviour as observed in optical conductivity spectra. 
At low temperature, the band structure is strongly modified, yielding to an insulating state with a optical gap of 120 meV. An accurate (3+1)D analysis of x-ray diffraction data shows that, although a phase transition does not occur, structural distortions increase as temperature is decreased, and vanadium clusterization is enhanced. We found that the changes of electronic properties and structure are intimately related. This indicates that structural distorsion play a major role in the insulating nature of (LaS)$_{1.196}$VS$_2$ and that electronic correlation may not be important, contrary to previous belief. These results shed a new light on the mechanism at the origin of non-linear electric properties observed in (LaS)$_{1.196}$VS$_2$.
\end{abstract}
\pacs{}
\maketitle
\section{Introduction}
The control of the electronic state of materials by an external perturbation has attracted considerable attention due to the possibility it offers to build novel electronic devices. One of the most famous examples is the colossal magnetoresistance observed in manganese perovskite oxides\cite{Kusters89,Jin94,Ramirez97}. In this case, the resistivity can decrease by orders of magnitude under magnetic field. So far most of the work have focussed on the effect of magnetic field or pressure\cite{Kuwahara95,Tomioka95,Tomioka96,Khazeni96,Kimura97,Moritomo97}. But recently another external perturbation, namely electric field, has attracted much attention, because of possible applications in the field of Resistive Random Access Memory. Indeed, numerous works have reported changes in resistivity of several orders of magnitude upon application of a strong electric field in transition metal oxides or chalcogenides\cite{Waser07,Asamitsu_97,Fors05,Kim06,Kozicki99,Beck00,Yamanouchi_99}. For example in Pr$_{0.7}$Ca$_{0.3}$MnO$_3$ \cite{Asamitsu_97} and La$_{2-x}$Sr$_x$NiO$_4$ \cite{Yamanouchi_99}, resistive switching were observed that may involve a dielectric breakdown of the charge ordered state, and collective depinning of holes on charge stripes, respectively. In the Mott insulator AM$_4$X$_8$ compounds, the electric field induced resistive switching is believed to result from a local Mott metal-insulator transition \cite{Vaju08, Vaju08bis, Dubost09, Cario10, Souchier11}. The underlying physical mechanisms of these phenomena are not yet fully understood, but all these studies point out that the mechanism is intimately related to the peculiar electronic ground state of these transition metal compounds. 

Recently, we have discovered that (LaS)$_{1.196}$VS$_2$ exhibits an electric field induced resistive switching at low temperature $\cite{Cario_2006}$. (LaS)$_{1.196}$VS$_2$ belongs to the series of misfit layered chalcogenides compounds of general formulae (LnX)$_{1+x}$TX$_2$ (Ln = rare earth ; X = S, Se ; T = Ti, V, Cr, Nb, Ta) \cite{Meerschaut_92}. As displayed in figure 1, the crystal structure of (LaS)$_{1.196}$VS$_2$ results in a regularly alternated stacking of rock salt type layers (LaS) and CdI$_2$ types layers (VS$_2$) along the $c$- axis \cite{Cario_2005}. Both VS$_2$ and LaS layers have different sub-cell parameters along the $a$ direction and their ratio (a(VS$_2$)/a(LaS)) is irrational. This mismatch between both layers is responsible for the incommensurability of (LaS)$_{1.196}$VS$_2$ along the $a$ direction and leads to a striking modulation of the vanadium atoms (see Fig. 1b). In (LaS)$_{1.196}$VS$_2$, a strong charge transfer exists from the LaS slab to the VS$_2$ slab leaving the $d$ bands of the vanadium atoms partially filled. This should confer a metallic character to (LaS)$_{1.196}$VS$_2$. However, so far, reports about the electronic state of (LaS)$_{1.196}$VS$_2$ are contradictory. Room temperature reflectivity measurements \cite{Kondo_95} and magnetic susceptibility measurements support a metallic character \cite{Cario_1999}. Conversely, transport measurements reveal a semiconductor-like behaviour \cite{Cario_1999}, and Photoemission spectroscopy suggest that (LaS$_2$)VS$_{1.19}$ is a strongly correlated system\cite{Imada98,Ino}. The nature of the ground state of (LaS)$_{1.196}$VS$_2$ still remains to be clarified which prevails the classification of (LaS)$_{1.196}$VS$_2$ among existing models of resistive switching i.e. breakdown of the charge ordered state, collective depinning of holes on charge stripes or local Mott Metal-insulator transition. In that respect a better understanding of the electronic ground state of (LaS)$_{1.196}$VS$_2$ is urgently needed.

The purpose of this paper is to investigate in details the structural and electronic properties of (LaS)$_{1.196}$VS$_2$ single crystals. We report optical and photoemission experiments that demonstrate that (LaS)$_{1.196}$VS$_2$ is neither a metal nor a correlated system, but behaves more like a band insulator whose band gap increases at low temperature. Our temperature X-ray diffraction experiments reveal that this band gap increase is related to the reinforcement of the Vanadium clusterization within the vanadium hexagonal VS$_2$ layer. Moreover, our work shed some light on the resistive switching found in (LaS)$_{1.196}$VS$_2$. It suggests that this phenomenon originates from a field assisted thermal weakening of the vanadium clusters and from the concomitant charge release.
\begin{figure}
\centering
\includegraphics[width=7cm]{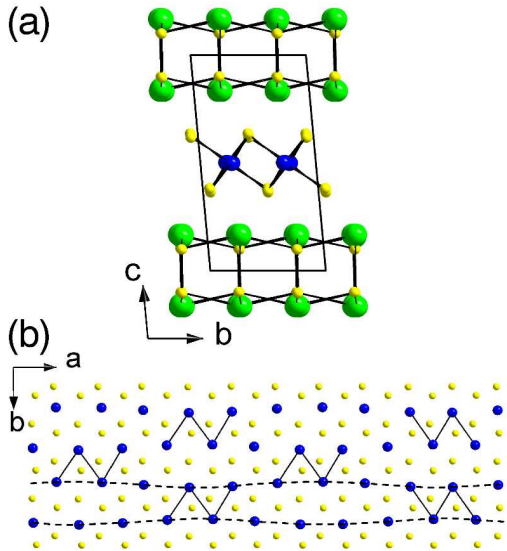}
\caption{Crystallographic structure of (LaS$_2$)VS$_{1.19}$.}
\label{Structure1}
\end{figure}
\section{Experimental details}
Pure powder of (LaS)$_{1.196}$VS$_2$ was obtained at 1200$^\circ$C from the sulfurization under H$_2$S gas of a mixture of binary oxides La$_2$O$_3$ and V$_2$O$_5$. Millimeter size crystals were grown by re-heating for ten days in a gradient furnace (950$^\circ$C-850$^\circ$C) the reaction-product combined with a small amount of iodine ($\leq$3 mg.cm$^{-3}$) to favor crystallization.

The (LaS)$_{1.196}$VS$_2$ crystals used for transport measurements were contacted using 50 $\mu$m gold wires and silver paste. The low-bias resistance of the (LaS)$_{1.196}$VS$_2$ was measured using a source-measure unit Keithley 236 by a standard four-probe technique. We checked that the contact resistances were much smaller than the sample resistance.

Single crystal diffraction experiments were performed on the same crystal thanks to a four circles FR 590 Nonius CAD-4F Kappa-CCD diffractometer between 90K and 300K, using Mo Kalpha radiation (0.071069 nm wavelength). All data treatments, refinement, and Fourier synthesis were carried out with the JANA2006 chain program \cite{JANA_2006}. A detailed procedure for data treatment and refinement can be found in \onlinecite{Cario_2005}. 

Angle-resolved photoemission spectra were recorded at the CASSIOPEE beamline of SOLEIL synchrotron with a SCIENTA-R4000 analyser and a total energy resolution of 15meV. Single crystals were cleaved in-situ and the absence of iodine (used during the synthesis) at the surface was checked through core-level analysis. Reproducible dispersions were obtained from 3 different samples. Charging effects were only observed below 20K. The reference Fermi level ($E_F$) was measured on the scraped copper plate onto which the samples were glued.

Near normal incidence reflectivity spectra were measured on 2*2 mm$^2$ $ab-$plane mirror-like surface of (LaS$_2$)VS$_{1.19}$ single crystals, using a BRUKER IFS 66v/S in the range 15 cm$^{-1}$- 55000 cm$^{-1}$, between 10K and 300K. After the initial measurement, the sample was coated {\it in situ} with a gold/aluminium film and remeasured at all temperatures. These additional data were used as reference mirrors to calculate the reflectivity in order to take into account light scattering on the surface of the sample. Optical conductivity spectra were obtained consistently by both Kramers-Kronig analysis and Drude-Lorentz fit procedure. 
\section{Results}
The electronic properties of (LaS)$_{1.196}$VS$_2$ are puzzling. Reported dc conductivity measurements \cite{Cario_1999} point out an insulating behaviour while magnetic susceptibility \cite{Cario_1999}, early photoemission \cite{Imada98,Ino} and optical conductivity \cite{Kondo_95} measurements performed on polycrystalline samples indicate a finite density of states at the Fermi level. In order to clarify the electronic ground state of (LaS)$_{1.196}$VS$_2$, optical and dc conductivity, as well as angle-resolved photoemission and structural measurements were reinvestigated as a function of temperature on large single crystals. 

First, several crystals were contacted with four electrodes to check the electrical behaviour. As shown in Fig.2, the resistivity exhibits a semiconductor-like behaviour and do not show any clear anomaly on the investigated temperature range. At low temperature, the resistiviy can be analyzed following a variable range hopping conduction mechanism given by :
\begin{equation}
\rho=\rho_0 e^{\left (\frac{T_0}{T}\right)^\alpha }
\end{equation}
where the exponent $\alpha$ can be obtained from the slope ($=-(1+\alpha)$) of $ ln \left( \frac{d ln \rho}{dT} \right) $ vs $ln(T)$. Below 50K, we found  $\alpha=0.52\pm 0.02$, which indicates that the resistivity is dominated by an Efros-Shklovskii variable range hopping mechanism (ES-VRH)\cite{Efros75}. Such behaviour usually highlights the important role of both disorder and long range coulomb interactions at low temperature. These results are fully consistent with prior measurements reported in the literature and therefore clearly attest of the quality of the crystals\cite{Cario_2006,Cario_1999}. 
\begin{figure}
\centering
\includegraphics[width=7cm]{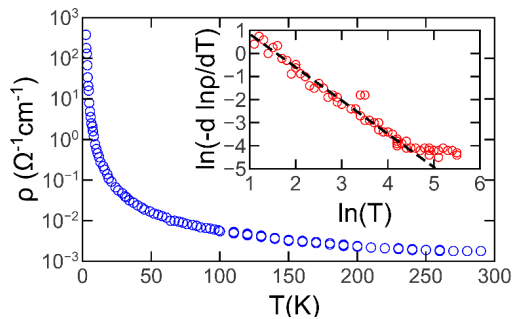}
\caption{Resistivity versus temperature measured on a (LaS$_2$)VS$_{1.19}$ single crystal. Inset : Plot of $ ln \left(-\frac{d ln \rho}{dT} \right) $ vs $ln(T)$. The dashed line is a linear fit of data at low temperature.}
\label{resistivity}
\end{figure}

\begin{figure}
\centering
\includegraphics[width=7cm]{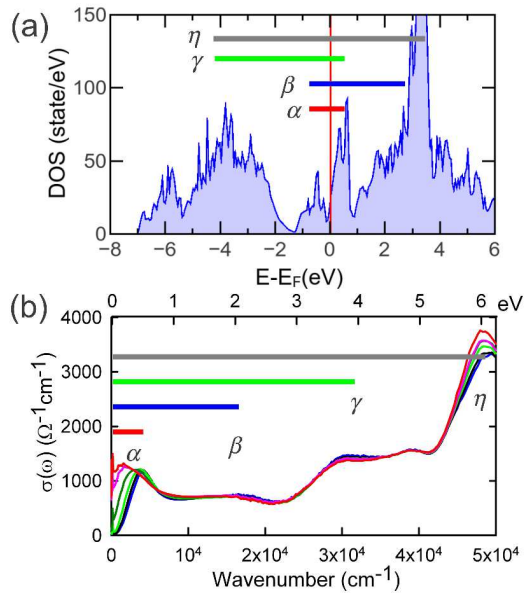}
\caption{(a) {\it ab-initio} LMTO-ASA band structure calculations. $\alpha$, $\beta$, $\gamma$, $\eta$ denote the optical transitions. (b)Optical conductivity up to 50000cm$^{-1}$ at 10 K, 100 K, 150 K, 200 K, 250 K and 300 K.}
\label{optics_1}
\end{figure}

We have subsequently performed optical conductivity measurements. Figure \ref{optics_1}(b) displays the optical conductivity spectra recorded from 300K to 10K on a freshly cleaved crystal of (LaS)$_{1.196}$VS$_2$. On a large energy scale a crude comparison of these data with {\it ab-initio} band structure calculations performed on the nearest commensurate supercell using LMTO-ASA method\cite{Cario_1999,Andersen84}, show a quite good agreement(Fig.\ref{optics_1}(a)). Indeed, excitations observed below 10000 cm$^{-1}$, at 16000 cm$^{-1}$, 30000 cm$^{-1}$ and 48000 cm$^{-1}$ are assigned to transitions within V $t_{2g}$ bands, V $t_{2g}\rightarrow$V $e_{g}$, S $3_{p}\rightarrow$V $t_{2g}$ and S $3_{p}\rightarrow$V $e_{g}$, respectively. 

Besides these general features, low energy electrodynamics exhibits a surprisingly strong temperature dependence. Above 200K, the optical spectra reveal a metallic-like character. Indeed, the far infrared conductivity is rather large, and extrapolation of the optical conductivity to $\omega \rightarrow 0$ reaches 800-900 $\Omega^{-1}$cm$^{-1}$, in fairly good agreement with the $\sigma_{dc}$ values. However, the optical conductivity exhibits a maximum at finite frequency (1500 cm$^{-1}$) and can not be successfully described by the usual Drude model. Such an anomalous shape of $\sigma(\omega)$ is usually attributed to disorder induced localization\cite{Mott85,Lee93,Tzamalis02}, strong electronic correlations\cite{Uchida91,Makino98,Baldassarre08,Basov11,Stewart12}or strong electron-lattice coupling(polarons)\cite{Holstein59,Bi93,Jung00,Jung01,Fratini06}. 
\begin{figure}
\centering
\includegraphics[width=8cm]{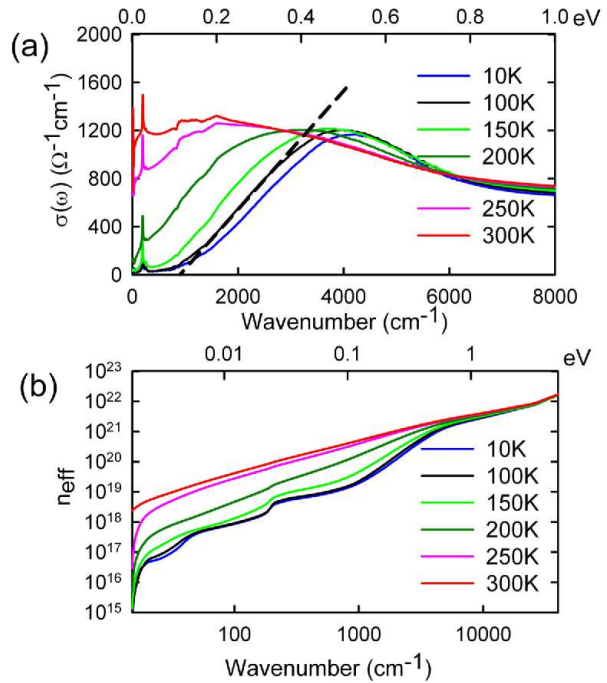}
\caption{(a) Optical conductivity in the far- and mid-infrared region. (b) Charge carrier effective density $n_eff(\omega _c)$ as function of the cut-off frequency $\omega_c$ at 10 K, 100 K, 150 K, 200 K, 250 K and 300 K.}
\label{optics_2}
\end{figure}

As temperature is decreased, a massive spectral weight transfer to high energy occurs, and a optical gap of 120 meV, estimated by extrapolating the steeply increasing part of $\sigma(\omega)$ to zero (dashed line in fig\ref{optics_2}(a), opens at 100 K. 
In order to quantify the modifications of low energy absorption, we define the effective density of charge carriers involved in the conductivity below $\omega_c$ : 
\begin{equation}
n_{eff} (\omega _c)=\frac{2 m_e}{\pi e^2} \int _0^{\omega_c}\sigma '(\omega)d \omega
\end{equation}
where $\omega_c$ is a cut-off frequency, $m_e$ is the bare electron mass. As shown in Fig.\ref{optics_2}, there is a considerable evolution at low frequencies. At 300K, choosing $\omega _c=$0.7 eV corresponding to the bandwidth found in ARPES(see below), we found $n_{eff}$=2.5.10$^{21}$ cm$^{-3}$, which is smaller than for a usual metal, but significantly larger than for a semiconductor. At low temperature, the far infrared spectral weight appears to be suppressed. According to the f-sum rule, $n_{eff} (\infty)$= const, the total area under $\sigma(\omega)$ must remain constant. Here, the spectral weight disappearing from the far infrared region is transferred to energies far above the gap value and the missing spectral weight is not recovered up to at least 1 eV (Fig.~\ref{optics_2}(b)). This implies that a  temperature change of 200K-300K produces a tremendous modification of the optical conductivity over an energy scale exceeding 1 eV (12000 K). 

On the one hand, such features (temperature dependence, spectral weight transfer) are hard to reconcile with a simple disorder induced localization picture. Actually, such unusual spectral weight transfer on large energy scale is known to occur in strongly correlated systems\cite{Rozenberg96,Georges96,Bluemer02,Baldassarre08,Basov11}. On the other hand, these results can however not be satisfactory explained by only invoking strong correlations, since the low frequency spectral weight is expected to grow at low temperature for doped Mott insulators, at variance with our observation.\cite{Jarrell95,Georges96,Taguchi02} Thus, neither disorder, nor electronic correlations alone can explain the puzzling properties of (LaS)$_{1.196}$VS$_2$. 

\begin{figure}
\centering
\includegraphics[width=7cm]{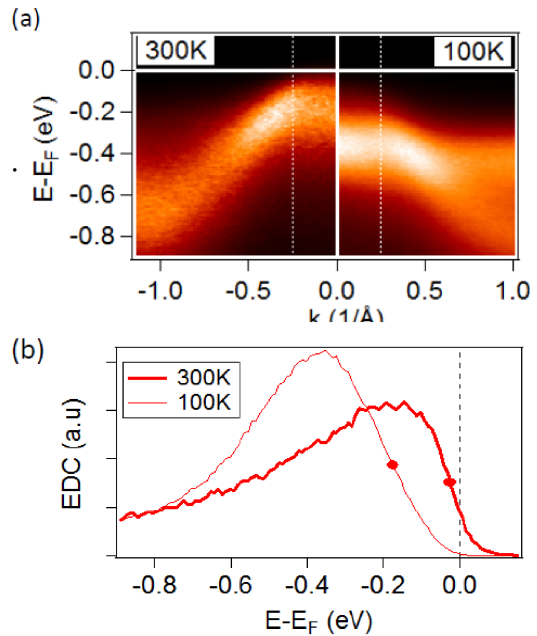}
\caption{(a) Energy-momentum images of the dispersion across the Zone center $\Gamma$ at 300K (left) and 100K (right), measured with a photon energy of 92 eV. (b) Energy distribution curves (EDC) along the dotted lines for the 2 temperatures.}
\label{ARPES}
\end{figure}

Angle resolved photoemission studies bring useful complementary light on this behavior. Three crystals of (LaS)$_{1.196}$VS$_2$, taken form the same batch were cleaved in vacuum to undertake these measurements. Figure \ref{ARPES}a displays representative energy-momentum intensity plots around the Brillouin Zone center $\Gamma$ (a more detailed analysis of the band structure will be given elsewhere \cite{Brouet_TR}). At 300K, a band with negative curvature approaches $E_F$ near k$_F$=0.2\AA$^{-1}$ (white dotted line) as if to form a small hole pocket around $\Gamma$. However, it does not really cross E$_F$, as shown by the small spectral weight present at E$_F$ in the energy distribution curve taken at this $k$ value (Fig. \ref{ARPES}b). This shows that the electronic structure is dominated by a nearly filled band, a situation where one would not expect a Mott insulator to form. On the other hand, the band dispersion is quite clear, ruling out a completely disordered case. These two findings fully support our previous conclusions from optical spectroscopy.  

When the temperature is lowered, the band shifts down almost uniformly by almost 100meV (Fig. \ref{ARPES}a), opening a much larger gap near k$_F$ (Fig. \ref{ARPES}b). Such an evolution will produce a redistribution of spectral weight over the energy scale of at least the entire band width (0.7eV), in good agreement with optical spectroscopy. If the mid-point of the leading edge is taken as a measure of this apparent gap, it increases from 20meV at 300K to 160meV at 100K. We note that 20meV is smaller than room temperature (25 meV), so that there will be many carrier thermally excited across the gap at this temperature. This high temperature pseudogap explains well the metallic-like spectra observed above 200K in optical spectroscopy. As the dispersion keeps a similar shape at high and low temperatures, we conclude that there is no dramatic reorganization of the electronic structure, but a gradual opening of a larger and larger gap. 

Both ARPES and optical conductivity measurements then suggest that (LaS)$_{1.196}$VS$_2$ is at the verge of a temperature induced insulator to metal transition, although electronic correlations may not be particularly strong. Moreover, the temperature dependence of the optical conductivity shares some similarity with that encountered in CDW, charge order or stripes systems such as Pr$_{1-x}$Ca$_x$MnO$_3$\cite{Okimoto99,Jung00} or La$_2$NiO$_{4+\delta}$\cite{Homes03,Poirot05}. Hence, the striking electronic properties of (LaS)$_{1.196}$VS$_2$ might be related to structural changes.
\begin{figure}
\centering
\includegraphics[width=7cm]{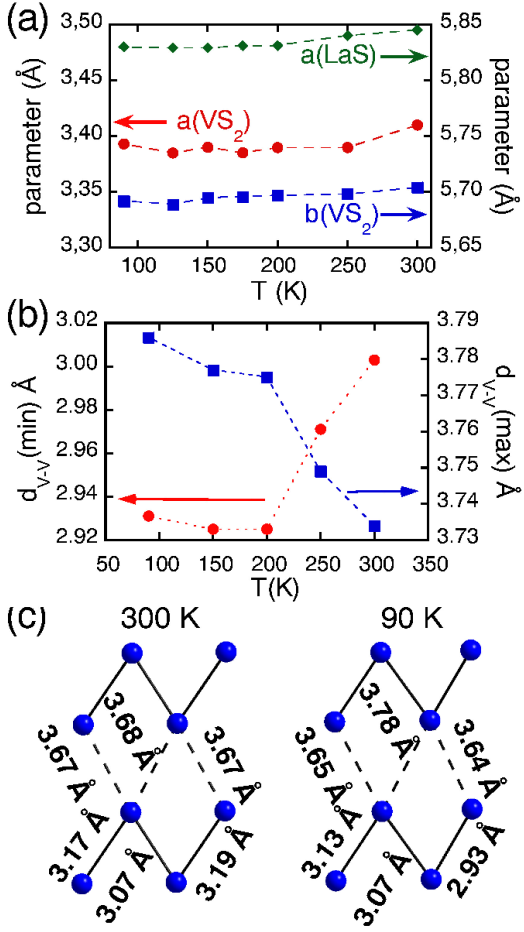}
\caption{ Evolution of the structure of (LaS)$_{1.196}$VS$_2$ between 300K and 90K (a) Temperature dependence of LaS and VS$_2$ layers in-plane cell parameters. (b) Temperature dependence of the shortest and the longest vanadium-vanadium distance $d_{minV-V}$ and $d_{maxV-V}$. (c) 
\textcolor{red}{Comparison of V-V distances observed within (plain lines) and between (dashed lines) some representatitve vanadium tetramers formed at 300 and 90K.} }
\label{structure2}
\end{figure}
A single crystal of (LaS)$_{1.196}$VS$_2$ was therefore mounted on a nonius CCD X-ray diffractometer and several data sets were collected down to 90K. At all temperatures the data were consistent with the triclinic symmetry and could be indexed in a (3+1)-D space using similar unit cell and modulation parameters as those used in our previous study at 300K (see reference \cite{Cario_2005} :  a$\simeq$3.41$\AA$, b$\simeq$5.84$\AA$, c$\simeq$11.19$\AA$, $\alpha\simeq$95.1$^\circ$, $\beta\simeq$84.8$^\circ$, $\gamma\simeq$90.0$^\circ$ and the following modulation vector q$\simeq$0.59 a* - 0.0 b*+ 0.0 c*). Figure \ref{structure2} presents the evolution of the $a$ and $b$ cell parameters between 300K and 90K. All parameters show smooth evolutions and no sign of structural phase transition was detected in the temperature range 90K-300K. Despite this lack of phase transition we have undertaken structural refinements of the low temperature data. The refinement procedure is described in detailed in our previous work focused on the room temperature data \cite{Cario_2005}. At 300K, the most striking feature of the (LaS)$_{1.196}$VS$_2$ structure is the large displacement modulation of the vanadium atoms along the $b$ direction. Figure 1(b) highlights that the atomic vanadium columns form along the $a$ direction sinusoidal-like waves with a strong modulation amplitude in the $b$ direction. 
\textcolor{red}{
Two adjacent vanadium atomic columns are out of phase which leads to short and long V-V distances where the atomic columns get closer and farther, respectively. In figure 1(b) the shortest V-V distances are drawn which emphasizes the formation of tetramers clusters with three short V-V distances. This figure shows also that two subsequent tetramer clusters along the b axis are separated by the longest V-V distances.}
Interestingly, our structural work demonstrates that the vanadium displacement modulation is strongly reinforced as the temperature decreases from 300K to 200K. Figure \ref{structure2} shows the longest and smallest V-V distances within the VS$_2$ layer as the function of the temperature. Clearly the shortest vanadium-vanadium distance drops and the longest vanadium-vanadium distance increases concomitantly below 200K. Figure  \ref{structure2}c compares some representative vanadium tetramer clusters formed at 300K and 90K (note that as the crystal is modulated the distances within the tetramers and between the tetramers vary from site to site). 
\textcolor{red}{At low temperature the vanadium clusterization is strongly reinforced, i.e. the vanadium tetramers get smaller, with one distance within the vanadium tetramer cluster dropping below 3Å. Concomitantly, some V-V distances separating the vanadium tetramer clusters (dashed line in figure 6c) get longer and can reach up to 3.78 Å.} Therefore, although a phase transition does not occur, our structural investigation of (LaS)$_{1.196}$VS$_2$ demonstrates that a strong clusterization reinforcement takes place in the vanadium atomic layer as the temperature decrease.
\section{Discussion}
\begin{figure}
\centering
\includegraphics[width=8cm]{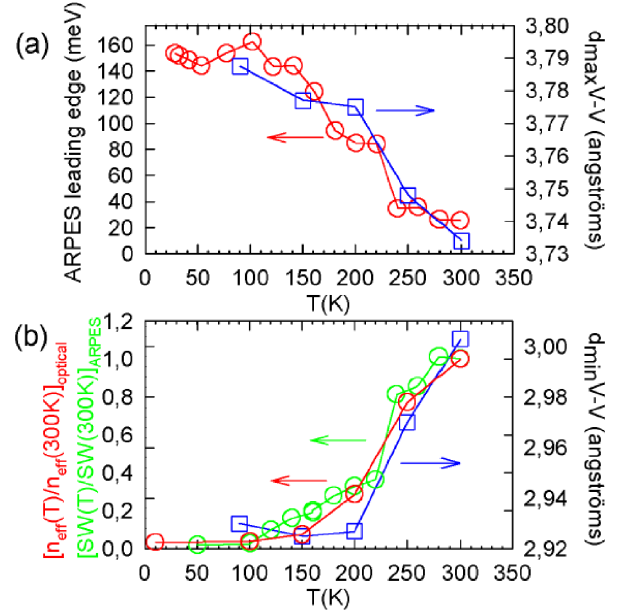}
\caption{(a) Temperature dependence of ARPES leading edge and longest vanadium-vanadium distance $d_{maxV-V}$. (b) Temperature dependence of effective charge carrier density at 120 meV, ARPES spectral weight at E$_F$ and shortest vanadium-vanadium distance $d_{minV-V}$. }
\label{finalccl}
\end{figure}
Interestingly optical and ARPES features (gap, spectral weight transfer) are intimately related to this clusterization reinforcement. The ARPES gap nicely scale with the largest V-V distance [Fig.~\ref{finalccl}(a)], while the spectral weight deduced again from both ARPES or optics scale with the smallest V-V distance [Fig. \ref{finalccl}(b)]. This strongly suggests that the evolution of the electronic structure is directly linked to the amplitude of the structural distortion. This behavior appears analogous to that of a Charge Density Wave(CDW) system\cite{Monceau, Monceau2} and suggests that a strong electron phonon coupling is at play in (LaS)$_{1.196}$VS$_2$. However, this misfit compound displays two noticeable differences compared to known CDW systems. First, the periodicity of the distortion is not connected to the Fermi Surface topology but imposed by the external potential of the LaS layer. Second, the amplitude of the distortion is not fixed to a particular value optimizing the energy balance between the cost of elastic energy and the gain of electronic energy, but strongly depends on temperature. A more adequate description is likely the clusterization picture depicted in Fig. \ref{structure2}(c). This will naturally create heterogeneous charge distributions with electron localization preferrentially on the vanadium clusters. The \lq\lq{degree of clusterization}\rq\rq{} seems to be the key parameter governing the conductivity. We have shown that temperature is obviously one parameter that can control the degree of metallicity, and it is natural to wonder whether other parameters should also tune this unusual insulating state\cite{Brouet_TR}.

These findings shed some light on the non linear electrical properties found in this system. First, it rules out a mechanism based on the breakdown of a Mott insulating state as previously proposed \cite{Cario_2006}. Conversely, it suggests that the non linear electrical properties observed in (LaS)$_{1.196}$VS$_2$ might share some common features with that encountered in CDW systems \cite{Monceau}. However the non linear properties associated \textcolor{red}{with} the depinning and collective motion of the CDW appear above a threshold field of the order of 0.01-0.1 V/cm \cite{Monceau2}. This is two order of magnitude lower than the threshold field observed in (LaS)$_{1.196}$VS$_2$ (around 50 V/cm) \cite{Cario_2006}. This discrepancy might be related to the strong pinning effect of the LaS layer. But, on the other hand, the nonlinearities in (LaS)$_{1.196}$VS$_2$ do not obey the Zener tunnelling law commonly observed in CDW systems \cite{Bardeen} or for the stripe depinning in La$_{2-x}$Sr$_x$NiO$_4$ \cite{Yamanouchi_99}. In that respect, the appearance of the non linear properties of (LaS)$_{1.196}$VS$_2$ are more likely related to the release of a large number of mobile charge carriers initially localized in the vanadium clusters than to a collective motion analogous to the sliding of CDW. Interestingly, the observation of a Poole-Frenkel mechanism (i.e., field assisted thermal emission of carriers) for the non linearity in (LaS)$_{1.196}$VS$_2$ \cite{Cario_2006} proves that the release of carriers is intimately related to the lowering of a thermal energy barrier by the electric field. we have observed that the electronic structure is very sensitive to changes of degree of vanadium clusterization with temperature. Our work therefore suggest that the electric field could have a similar effect than temperature. This effetc could weaken the vanadium clusterization and subsequently reduce the localisation strengh on these clusters. Such a mechanism would cause the breakdown of the insulating state and therefore easily explain the non linear electrical properties of (LaS)$_{1.196}$VS$_2$.  
\section{Conclusion}
Transport, ARPES, optical spectroscopy and x-ray diffraction were combined to investigate in details the electronic structure of (LaS)$_{1.196}$VS$_2$. We found that, unlike claimed in previous studies, (LaS)$_{1.196}$VS$_2$ is neither a metal, nor a strongly correlated system, but behaves like a narrow gap insulator at the verge of an insulator to metal transition. 
The  gap is small enough to produce pseudo-metallic behavior at 300K. At low temperature, band structure is dramatically modified, although no structural transition occurs. Concomitantly, an optical gap of 120 meV opens. Such effects are explained by unusually large structural changes, {\it i.e.} vanadium clusterization, due to the peculiar incommensurate structure of (LaS)$_{1.196}$VS$_2$. As a consequence, our results suggest that non linear electrical properties of (LaS)$_{1.196}$VS$_2$ are related to field assisted vanadium cluster reorganization.

This work was supported by the French Agence Nationale de la Recherche through the funding of the "NanoMott" (ANR-09-Blan-0154-01) project.


\begin{thebibliography}{}
\bibitem{Kusters89} R.M. Kusters, J. Singleton, D.A. Keen, R. McGreevy, W. Hayes, Physica (Amsterdam), 155B :362, 1989.
\bibitem{Jin94} S. Jin, T.H. Tiefel, M. McCormack, R. A. Fasnacht, R. Ramesh, and L. H. Chen. Science, 264 :413, 1994.
\bibitem{Ramirez97}A. P. Ramirez, J. Phys.: Condens. Matter 9, 8171 (1997)
\bibitem{Kuwahara95}H. Kuwahara, Y. Tomioka, A. Asamitsu, Y. Moritomo, Y. Tokura, Science 270, 961 (1995). 
\bibitem{Tomioka95}Y. Tomioka, A. Asamitsu, Y. Moritomo, H. Kuwahara, and Y. Tokura, Phys. Rev. Lett. 74, 5108 (1995)
\bibitem{Tomioka96}Y. Tomioka, A. Asamitsu, H. Kuwahara, Y. Moritomo, Y. Tokura, Phys. Rev. B 53, R1689 (1996)
\bibitem{Khazeni96}K. Khazeni, Y. X. Jia, Li Lu, Vincent H. Crespi, Marvin L. Cohen, and A. Zettl, Phys. Rev. Lett. 76, 295 (1996)
\bibitem{Kimura97}T. Kimura, A. Asamitsu, Y. Tomioka and Y. Tokura, Phys. Rev. Lett. 79, 3720 (1997)
\bibitem{Moritomo97}Y. Moritomo, H. Kuwahara, and Y. Tomioka and Y. Tokura, Phys. Rev. B 55, 7549 (1997)
\bibitem{Waser07}R. Waser, M. Aono, Nature Materials 6, 833(2007). 
\bibitem{Asamitsu_97} A. Asamitsu, Y. Tomioka, H. Kuwahara and Y. Tokura, Nature \textbf{388}, 50 (1997). 
\bibitem{Fors05}R. Fors, S. I. Khartsev, A. M. Grishin, Phys. Rev. B 71, 045305 (2005). 
\bibitem{Kim06}D. S. Kim, Y. H. Kim, C. E. Lee, Y. T. Kim,  Phys. Rev. B 74, 174430 (2006).
\bibitem{Kozicki99}M. N. Kozicki, M. Yun, L. Hilt and A. Singh , Pennington NJ USA: Electrochem. Soc. 298 (1999)
\bibitem{Beck00}A. Beck, J. G. Bednorz,C. Gerber, C. Rossel, D. Widmer, Appl. Phys. Lett. 77, 139 (2000)
\bibitem{Yamanouchi_99} S. Yamanouchi, Y. Taguchi and Y. Tokura, Phys. Rev. Lett. \textbf{83}, 5555 (1999); V. Sachan, D. J. Buttrey, J. M. Tranquada, J. E. Lorenzo and G. Shirane, Phys. Rev. B \textbf{51}, 12742 (1995).
\bibitem{Vaju08} C. Vaju, L. Cario, B. Corraze, E. Janod, V. Dubost, T. Cren, D. Roditchev, D. Braithwaite, O. Chauvet, Advanced Materials \textbf{20}, 2760 (2008).
\bibitem{Vaju08bis} C. Vaju, L. Cario, B. Corraze, E. Janod, V. Dubost, T. Cren, D. Roditchev, D. Braithwaite and O. Chauvet, Microelectronics Engineering \textbf{85}, 2430 (2008).
\bibitem{Dubost09} V. Dubost, T. Cren, C. Vaju, L. Cario, B. Corraze, E. Janod, F. Debontridder, D. Roditchev, Advanced Functional Materials \textbf{19}, 2800 (2009).
\bibitem{Cario10} L. Cario, C. Vaju, B. Corraze, V. Guiot, E. Janod, Advanced Materials \textbf{22}, 5193 (2010).
\bibitem{Souchier11} E. Souchier, L. Cario, B. Corraze, P. Moreau, P. Mazoyer, C. Estouns, R. Retoux, E. Janod, and M.-P. Besland, physica status solidi (RRL) - Rapid Research Letters \textbf{5}, 53 (2011).
\bibitem{Cario_2006} L. Cario, B. Corraze, A. Meerschaut, O. Chauvet, Phys. Rev. B \textbf{73}, 155116 (2006).
\bibitem{Meerschaut_92} A. Meerschaut (ed.), \textit{Incommensurate Sandwiched Layered Compounds}, Materials Science Forum, 100-101 (1992).
\bibitem{Cario_2005} L. Cario \textit{et al.}  Materials Research Bulletin \textbf{40}, 125 (2005).
\bibitem{Kondo_95} T. Kondo, K. Suzuki, and T. Enoki, J. Phys. Soc. Jpn \textbf{64}, 4296 (1995); T. Kondo, K. Suzuki, T. Enoki, H. Tajima and T. Ohta,  J. Phys. Chem. Sol. \textbf{57}, 1105 (1996).
\bibitem{Cario_1999} L. Cario, J. Rouxel, A. Meerschaut, Y. Moelo, B. Corraze and O. Chauvet, J. Phys. Cond. Matter, \textbf{11} 2887 (1999); L. Cario, B. Corraze, A. Meerschaut, Y. Moelo and O. Chauvet, Synth. Met. \textbf{103}, 2640 (1999).
\bibitem{Imada98} M. Imada, A. Fujimori, Y. Tokura , Rev. Mod. Phys. 70, 1039–1263 (1998)
\bibitem{Ino} A. Ino \textit{et al.} Phys. Rev. B \textbf{69}, 195116 (2004).
\bibitem{JANA_2006} V. Petricek, M. Dusek, L. Palatinus, JANA 2006, Academy of Science of Cseck Republic: Praha, 2006.
\bibitem{Efros75} A. L. Efros and B. I. Shklovskii, J. Phys. C 8, 249 (1975).
\bibitem{Andersen84} O. K. Andersen and O. Jepsen, Phys. Rev. Lett. 53, 2571 (1984)
\bibitem{Mott85}N. F. Mott and M. Kaveh, Adv. Phys. 34, 329 (1985)
\bibitem{Lee93}K. Lee, A. J. Heeger, Y. Cao, Phys. Rev. B 48, 14884 (1993)
\bibitem{Tzamalis02}G. Tzamalis, N. A. Zaidi, C. C. Homes and A. P. Monkman, Phys. Rev. B 66, 085202 (2002)
\bibitem{Uchida91}S. Uchida, T. Ido, H. Takagi, T. Arima, Y. Tokura, and S. Tajima, Phys. Rev. B 43, 7942(1991).
\bibitem{Makino98} H. Makino, I. H. Inoue, M. J. Rozenberg, I. Hase, Y. Aiura, S. Onari, Phys. Rev. B 58, 4384 (1998)
\bibitem{Baldassarre08} L. Baldassarre, A. Perucchi, D. Nicoletti, A. Toschi, G. Sangiovanni, K. Held, M. Capone, M. Ortolani, L. Malavasi, M. Marsi, P. Metcalf, P. Postorino, S. Lupi, Phys. Rev. B 77, 113107 (2008)
\bibitem{Basov11} D. N. Basov, Richard D. Averitt, Dirk van der Marel, Martin Dressel, and Kristjan Haule, Rev. Mod. Phys. 83, 471 (2011). 
\bibitem{Stewart12} M. K. Stewart, D. Brownstead, S. Wang, K. G. West, J. G. Ramirez, M. M. Qazilbash, N. B. Perkins, I. K. Schuller, and D. N. Basov, Phys. Rev. B 85, 205113 (2012)
\bibitem{Holstein59}T. Holstein, Ann. Phys. (N.Y.) 8, 325(1959).
\bibitem{Bi93}X.-X.Bi, P. C. Eklund, and J. M. Honig, Phys. Rev. B 48,3470(1993)
\bibitem{Jung00}J. H. Jung, H. J. Lee, T. W. Noh, E. J. Choi, Y. Moritomo, Y. J. Wang, and X. Wei,Phys. Rev. B 62, 481(2000)
\bibitem{Jung01}J. H. Jung, D.-W. Kim, T. W. Noh, H. C. Kim, H.-C. Ri, S. J. Levett,M. R. Lees, D. M. Paul, and G. Balakrishnan, Phys. Rev. B 64, 165106(2001).
\bibitem{Fratini06}S. Fratini, S. Ciuchi, Phys. Rev. B 74, 075101 (2006)
\bibitem{Rozenberg96}M. J. Rozenberg, G. Kotliar and H. Kajueter, Phys. Rev. B 54, 8452 (1996) 
\bibitem{Georges96} A. Georges, G. Kotliar, W. Krauth and M. Rozenberg, Rev. Mod. Phys. 68, 13 (1996)
\bibitem{Bluemer02}N. Bl\"umer,  Ph.D. thesis (Universit\"at Augsburg)(2002)
\bibitem{Jarrell95}M. Jarrell,J. K. Freericks,Th. Pruschke, Phys. Rev. B 51, 11704 (1995) 
\bibitem{Taguchi02}Y. Taguchi, K. Ohgushi, and Y. Tokura, Phys. Rev. B 65, 115102 (2002) 
\bibitem{Brouet_TR} V. Brouet \textit{et al.}, in preparation (to appear on condmat).
\bibitem{Monceau}  P. Monceau, \textit{Electronic properties of Inorganic Quasi-One-Dimensional Materials, II},( D. Riedel Publishing Company 1985).
\bibitem{Monceau2}  P. Monceau, et al. Phys. Rev. Let.  37, 602 (1976)
\bibitem{Bardeen}  J. Bardeen,  Phys. Rev. Let.  45, 1978(1980)
\bibitem{Okimoto99}Y. Okimoto, Y. Tomioka, Y. Onose, Y. Otsuka, Y. Tokura, Phys. Rev. B 59, 7401-7408 (1999)
\bibitem{Homes03}C. C. Homes, J. M. Tranquada, Q. Li, A. R. Moodenbaugh, and D. J. Buttrey, Phys. Rev. B 67, 184516(2003).
\bibitem{Poirot05}N. Poirot, V. Ta Phuoc, G. Gruener, and F. Gervais, Solid State Sci. 7, 1157 (2005)
\end{thebibliography}
\end{document}